\documentclass[fleqn,usenatbib]{mnras}

\usepackage{newtxtext,newtxmath}

\usepackage[T1]{fontenc}
\usepackage{orcidlink}
\DeclareRobustCommand{\VAN}[3]{#2}
\let\VANthebibliography\thebibliography
\def\thebibliography{\DeclareRobustCommand{\VAN}[3]{##3}\VANthebibliography}

\usepackage{graphicx}	
\usepackage{amsmath}	
\usepackage{xcolor}

\title[The \ion{H}{i} of LCBGs in CHILES]{CHILES XII: The \ion{H}{i} evolution of Luminous Compact Blue Galaxies between $0<z<0.48$}

\author[H. Arlow et al.]{
Henco Arlow\textsuperscript{\orcidlink{0000-0002-5815-9407},1}\thanks{E-mail: hencoarlow@gmail.com}
D. J. Pisano\textsuperscript{\orcidlink{0000-0001-7996-7860},1}
Matthew A. Bershady\textsuperscript{\orcidlink{0000-0002-3131-4374},2}
Lucas R. Hunt\textsuperscript{\orcidlink{0000-0001-8587-9285},3}
Nicholas Luber\textsuperscript{\orcidlink{0009-0006-6641-0928},4}
\newauthor
Jennifer Donovan Meyer\textsuperscript{\orcidlink{0000-0002-3106-7676},5}
Emmanuel Momjian\textsuperscript{\orcidlink{0000-0003-3168-5922},3}
Julia Blue Bird\textsuperscript{\orcidlink{0000-0002-9627-7519},3}
and Hansung B. Gim\textsuperscript{\orcidlink{0000-0003-1436-7658},6}
\\
$^{1}$Department of Astronomy, University of Cape Town, Private Bag X3, Rondebosch, 7701 Cape Town, Republic of South Africa\\
$^{2}$University of Wisconsin, Department of Astronomy, 475 N. Charter St., Madison, WI 53706, USA\\
$^{3}$National Radio Astronomy Observatory, P.O. Box O, Socorro, NM 87801, USA\\
$^{4}$Department of Astronomy, Columbia University, Mail Code 5247, 538 West 120th Street, New York, NY 10027, USA\\
$^{5}$National Radio Astronomy Observatory (NRAO), 520 Edgemont Road, Charlottesville, VA 22903, USA\\
$^{6}$Eureka Scientific Inc., 2452 Delmer Street, Suite 100, Oakland, CA 94602\\
}

\date{Accepted 2026 April 14. Received 2026 April 14; in original form 2026 February 24}

\pubyear{2026}

\begin{document}
\label{firstpage}
\pagerange{\pageref{firstpage}--\pageref{lastpage}}
\maketitle

\begin{abstract}
We study the evolution of Luminous Compact Blue Galaxies (LCBGs) by making use of \ion{H}{i} emission line data provided by the full 856\,h COSMOS \ion{H}{i} Large Extragalactic Survey (CHILES), which spans a redshift range of $0\leq z \leq 0.48$ within the COSMOS field. We report the results on a cubelet stacking analysis, which we use to estimate the average \ion{H}{i} mass evolution of LCBGs in the field up to $z=0.48$. For the stacks that do not show a detection, we report an upper limit estimate of the average \ion{H}{i} mass. We also report on two directly detected LCBGs. We find the average \ion{H}{i} mass in LCBGs at redshifts $z=0.26$, $z=0.35$ and $z=0.45$ respectively to be $\langle M_{\rm HI}\rangle<4.89\times10^9$\,M$_\odot$, $\langle M_{\rm HI}\rangle=(2.49\pm0.75)\times10^9$\,M$_\odot$ and $\langle M_{\rm HI}\rangle=(6.44\pm2.71)\times10^9$\,M$_\odot$. We see no strong evidence for evolution in the average \ion{H}{i} mass over this redshift range, consistent with other recent studies of the evolution of the \ion{H}{i} in galaxies at $z<0.5$. On average, LCBGs appear to retain substantial gas reservoirs, with gas fractions staying constant and remaining broadly consistent with those of the larger star-forming population. LCBG gas depletion timescales are nearly an order of magnitude shorter than in normal star-forming galaxies across the studied redshift range, aligning with the period during which their number density drops sharply.
\end{abstract}




\section{Introduction}
Rapidly evolving Luminous Compact Blue Galaxies (LCBGs), which are heterogeneous in terms of their morphology and star formation properties, provide a way to study galaxy evolution more broadly. These galaxies comprise a subset of starburst galaxies and are abundant at higher redshifts ($z\sim1$), but rare locally \citep{Guzman:1997, Werk:2004,Hunt:2021}, and were first identified from sources that showed emission lines consistent with star formation in surveys looking for faint quasi-stellar objects by \citet{Koo:1994}. These galaxies are traditionally defined according to the following criteria: they have a rest-frame total absolute magnitude of  $M_B<-18.5$\,mag, a colour of $B-V<0.6$\,mag and an effective surface brightness of $SB_e(B)<21$\,mag\,arcsec$^{-2}$ \citep{Werk:2004}. Local LCBGs tend to be small in physical size with their effective radii ($R_e$) being $<3$\,kpc \citep{Garland:2004}. Their compact classification reflects their high surface brightness and small effective radii relative to their luminosity. Some studies split LCBGs into two different classifications. Around ~60\% of the sample in a study by \citet{Guzman:1997} at $0.1<z<1.4$ were classified as \ion{H}{ii}-like galaxies, where \ion{H}{ii} denotes ionised hydrogen. These systems share properties with local \ion{H}{ii} galaxies, which are luminous, star-bursting dwarfs galaxies with spectra similar to galactic \ion{H}{ii} regions. The remaining ~40\% of the sample were classified as SB disc-like. These systems have different star formation histories, with much lower star formation rates (SFRs) locally than in the past. They show similarities to local spiral galaxies and giant irregular galaxies. A comparison can be drawn between LCBGs and Green Pea (GP) galaxies in terms of blue luminosity, compactness, metallicity, and morphology. While LCBGs are defined photometrically through their high blue luminosities, blue colours, and compact sizes, Green Pea galaxies are identified spectroscopically by their extremely strong optical emission lines, particularly [\ion{O}{iii}] $\lambda5007$ \citep{Cardamone:2009}. As a result, GPs typically occupy the lower-luminosity and lower-mass end of the LCBG parameter space, while often being similarly compact and comparably blue due to intense, compact starburst activity. Green Peas have lower redshift counterparts known as Blueberry galaxies \citep{Yang:2017}, which are dwarf starbursts that are also similar to LCBGs. LCBGs have been suggested to represent lower-redshift, lower-mass counterparts to Lyman Break Galaxies as well \citep{Hoyos:2004} and there are further overlaps with Ultra Luminous Infrared Galaxies \citep{Heckman:2005}. Finally we note that LCBGs are distinct from Blue Compact Dwarf galaxies, being systematically more massive, more luminous, and more metal-rich \citep{Garland:2004, Tollerud:2010}.

The physical processes that trigger their star-busting nature remain an open question. While it is unclear whether internal mechanisms play a significant role, external interactions such as encounters with nearby galaxies could be one of the drivers. The presence of companions appears to strongly influence the star-forming activity in these galaxies. Around 50\% of a local sample of 29 LCBGs studied by \citet{Garland:2015} ($z \sim 0$) had one or more optical companions within 100 kpc, and \citet{Gallego:Thesis} found companions in 43\% of their sample of 22 LCBGs ($0.007<z<0.042$). Furthermore, 20\% of LCBGs in the study of \citet{Garland:2015} showed signs of being the result of a merger, while \citet{Gallego:Thesis} reported that 10\% had kinematics consistent with minor mergers and 5\% showed evidence of major mergers. These interactions likely contribute to the heterogeneous morphologies observed in LCBGs. They are generally irregular or possess spiral structures, with local examples often exhibiting clumpy features typical of disturbed star-forming regions. Such morphological variety is consistent with the effects of gravitational interactions and merging activity, as commonly seen in starburst galaxies.
 
It is likely that these interactions are driving the rapid star formation and hence their rapid evolution. LCBGs contribute around 45\% of the fractional increase in the SFR density between $0.4<z<1$ \citep{Guzman:1997}. Their total SFR seems to increase with redshift, as observed by \citet{Hunt:2021}, with the average of their sample's SFRs at $z=0.92$ ($\langle \mathrm{SFR} \rangle = 53.3 \pm 27.6 \,\mathrm{M_\odot\,yr^{-1}}$) being approximately 25 times higher than those at $z=0.12$ ($\langle \mathrm{SFR} \rangle = 1.99 \pm 0.73 \,\mathrm{M_\odot\,yr^{-1}}$). Using deep field data from the GAMA region within the COSMOS field, \citet{Hunt:2021} demonstrated that from $z\sim0.1$ to $z\sim1$ the characteristic luminosity of LCBGs increases by $\sim0.22$ mag, accompanied by a fourfold increase in number density. At $z=0.9$, they found that LCBGs account for approximately 50\% of the luminosity density of galaxies with $M_B < -18.5$\,mag. With less than 2\% of the galaxies in the local universe being LCBGs, they are clearly a very rapidly evolving class of galaxies. The main challenge has been trying to identify what they evolve into.

The evolutionary trajectory of LCBGs is still one of the main unanswered questions. In terms of co-moving number density, LCBGs evolve at a faster rate than spirals, but not as fast as irregular galaxies \citep{Hunt:2021}. There are two schools of thought when it comes to the evolutionary end state of LCBGs. One theory is that they may be the progenitors of the bulges of massive local spiral galaxies \citep{Phillips:1997,Hammer:2001}, which would imply the presence of substantial neutral gas reservoirs capable of sustaining ongoing star formation and subsequent disk growth. Another theory is that they evolve into lower mass elliptical galaxies, because the luminosities and surface brightness of \ion{H}{ii}-like LCBGs are likely to fade to be comparable to dwarf ellipticals \citep{Koo:1994}. Dwarf ellipticals (dEs) are smaller, less luminous elliptical galaxies. This scenario would be favoured if LCBGs exhibit low \ion{H}{i} gas fractions or short gas depletion timescales, consistent with rapid gas exhaustion or removal. 

If we wish to understand the galactic evolution of galaxies between $0<z<1$, LCBGs offer a great laboratory because of their rapidly evolving nature. Neutral hydrogen, \ion{H}{i}, is the most direct tracer of a galaxy’s fuel reservoir, yet detecting it beyond the local Universe is notoriously difficult because the \ion{H}{i} line rapidly weakens with redshift. Until recently, meaningful \ion{H}{i} constraints at intermediate redshifts were almost entirely out of reach. With modern interferometers, however, we can finally begin to probe the cold gas content of LCBGs during the epoch when their evolution is most rapid. This opens a new window onto how these compact, intensely star-forming systems acquire, deplete, and cycle their gas.

For all calculations in this paper, we assume $H_0=70$\,km\,s$^{-1}$\,Mpc$^{-1}$, $\Omega_M=0.3$ and $\Omega_{\Lambda}=0.7$ in a flat $\Lambda$CDM model \citep{Planck}.

\begin{figure*}
\includegraphics[width=0.9\textwidth]{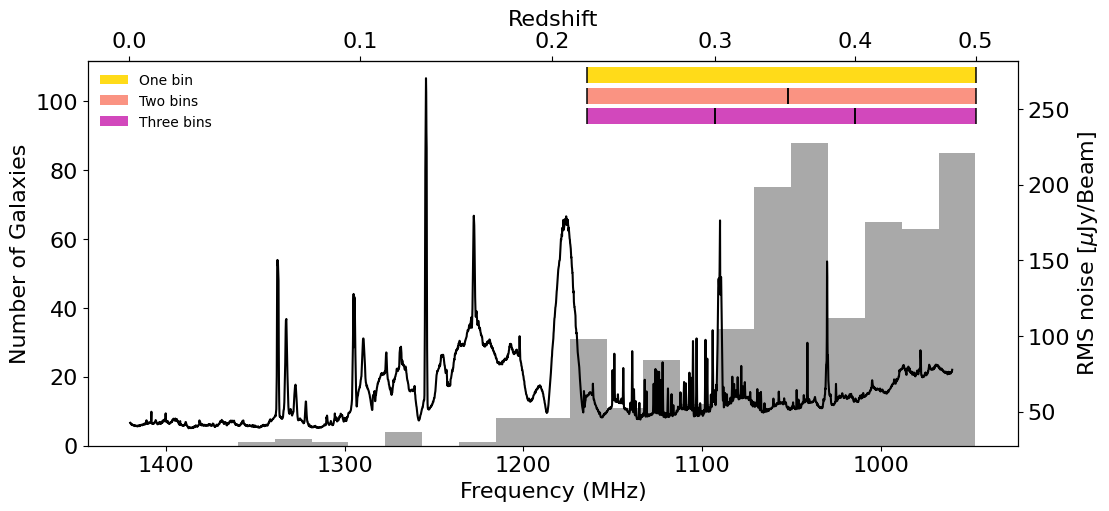}
\caption{The RMS noise per channel (in black, right axis) over the full frequency range of the CHILES cube. The histogram (in grey, left axis) indicates the distribution of LCBGs in our sample, binned according to the frequency at which we expect to detect the \ion{H}{i} 21\,cm line.}
\label{fig:CHILES_noise}
\end{figure*}

\begin{figure}
\includegraphics[width=\columnwidth]{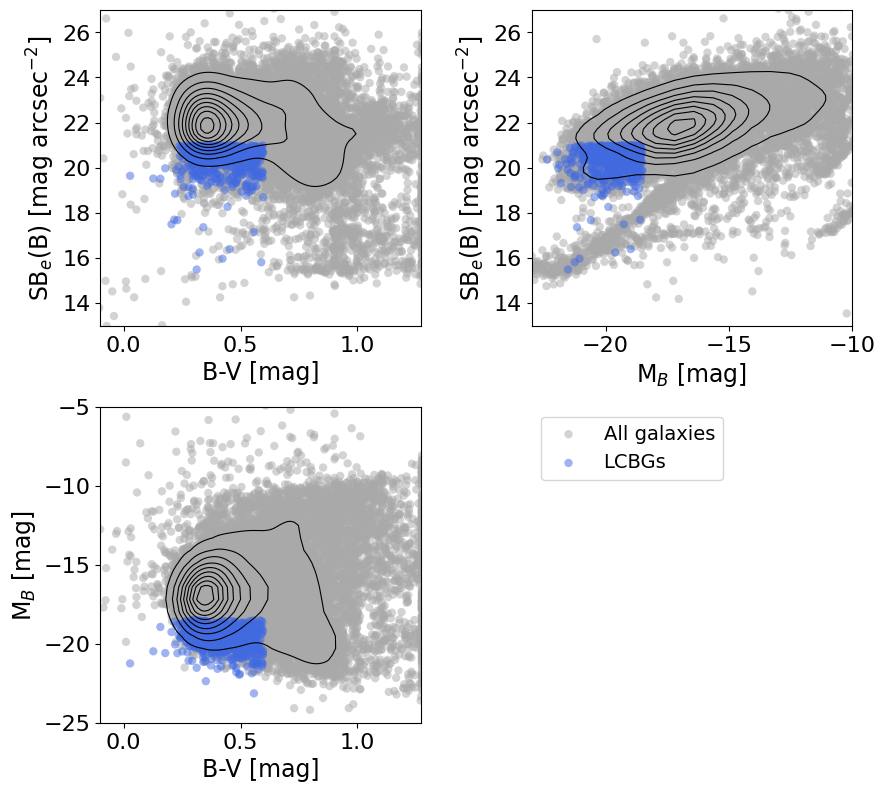}
\caption{Our full catalog of 552 LCBGs (blue) identified from all galaxies in the G10 region (grey) via their color ($B-V$), B-band magnitude ($M_B$) and surface brightness ($SB_e(B)$). All galaxies represented here have spectroscopic redshifts < 0.48. We see the LCBGs populating the one corner of the parameter space. In each sub-figure, the grey points in the LCBG quadrant represent those galaxies that do not meet the third criteria in each case. The contours represent the number density of galaxies in the parameter space.}
\label{fig:cutoffs}
\end{figure}

\section{Data}
The COSMOS \ion{H}{i} Large Extragalactic Survey, CHILES, is a neutral hydrogen deep field survey done by the Karl G. Jansky Very Large Array in its B-configuration. Observations ran over five epochs from 2013 to 2019. The survey, situated in the COSMOS field, is centered at $(10^{h}01^{m}24^{s}, +02^{\circ}21'00'')$ and covers a redshift range of $0<z<0.48$ (960 to 1420\,MHz) making it one of the first surveys to continuously observe \ion{H}{i} in this redshift range \citep{Hess:2019,BlueBird:2020,Luber:2025a}. COSMOS is a very well studied patch of sky, rich with ancillary data spanning wavelengths from X-ray to radio \citep{Koekemoer:2007}. The region is devoid of gas from the Milky Way and contains minimal stars and strong radio continuum sources, i.e. radio galaxies or supernova remnants.

\citet{Hess:2019} demonstrated that CHILES can yield high-quality data even in redshift ranges heavily effected by radio frequency interference (RFI) due to satellites or ground based radar. They detected 16 galaxies and analysed their environmental properties, yielding results consistent with established scaling relations \citep{BlueBird:2020}. This study utilised the full dataset. The full survey is comprised of 856\,h of observations done over several years, split up into five epochs. The final epoch was completed in 2019. \citet{Luber:2025a} presented the first full-epoch analysis of CHILES, using stacking out to $z = 0.48$ to measure average \ion{H}{i} masses in blue cloud and red sequence galaxies. In their blue sample, they found that while intermediate-stellar-mass galaxies ($10^{9-10}$\,M$_\odot$) maintain a constant \ion{H}{i} content over time, high-stellar-mass galaxies ($10^{10-12.5}$\,M$_\odot$) become increasingly gas-poor with decreasing redshift, suggesting distinct evolutionary pathways within the star-forming population. \citet{Luber:2025b} further explored stacking in the data to describe the redshift evolution of \ion{H}{i} of blue galaxies in different cosmic web environments. They observe a decrease in \ion{H}{i} gas fraction with increasing distance from filaments. Resolved \ion{H}{i} emission has been directly detected in the data up to a redshift of 0.47 \citep{BlueBird:2026}.

The main properties of the final CHILES cube are shown in Table~\ref{tab:TableCHILES} (as described in  \citet{Luber:2025a}). The cube was produced with a pixel scale of $2''\times2''$ and a spectral resolution of 0.125\,MHz which corresponds to $\approx 26.4$\,km/s wide channels at $z=0$ and $\approx 39.6$\,km/s wide channels at $z=0.48$. The cube is $40'\times 40'$ in size with a synthesised beam which was smoothed to a common resolution of $9''\times9''$. The cube has an average noise of 63.5\,$\mu$Jy\,beam$^{-1}$ per channel and a mode of 39.0\,$\mu$Jy\,beam$^{-1}$, but we see a general increase in noise due to ground based radars between 1160\,MHz and 1300\,MHz (see Figure~\ref{fig:CHILES_noise}). We also observe spikes in noise at around 1030\,MHz, 1090\,MHz and 1333\,MHz.

\begin{table}
\centering
  \caption{Summary of properties of our final CHILES cube.}
  \label{tab:TableCHILES}
  \begin{tabular}{lr} 
  \hline
   Pointing & $10^{h}01^{m}24^{s} +02^{\circ}21'00''$ \\
   Frequency range & 960\,MHz - 1420\,MHz\\
   Pixel size & 2$''$\\
   Image size & $1200\times1200$\,pixels\\
   Integration Time &  856\,h\\
   Observations & 5/10/2013 - 11/04/2019\\
   Spectral Resolution & 125\,kHz\\
   Velocity Resolution & 26.4\,km/s ($z=0$) - 39.6\,km/s ($z=0.48$)\\
   Synthesized Beam (FWHM) & $9''\times9''$\\
   \hline
   \end{tabular}
\end{table}

Many surveys have taken place within the COSMOS region, making it a very well studied patch of sky. The LCBGs in our study were originally identified within the COSMOS survey, with the relevant photometric and spectroscopic data reprocessed as part of the Galaxy and Mass Assembly (GAMA) survey \citep{Driver:2011}. The survey covers many areas of the sky. The part of the survey that is centered in COSMOS is labeled as G10 and consists of around 18000 sources containing photometric data from UV to FIR \citep{Davies:2015a}. The three main identifiers of LCBGs (i.e., the rest-frame absolute magnitude in the B-band, colour and surface brightness) were calculated by \citet{Hunt:2021}. We use this selection of LCBGs for our analysis. Figure~\ref{fig:cutoffs} shows our full sample of 552 LCBGs and how it fits within this parameter space. The spectroscopic redshifts we use, which are generated using spectra from the zCOSMOS survey and reprocessed by the GAMA team \citep{Davies:2015b}, are also from the dataset published by \citet{Hunt:2021}.

\section{Stacking}

\begin{figure}
\includegraphics[width=\columnwidth]{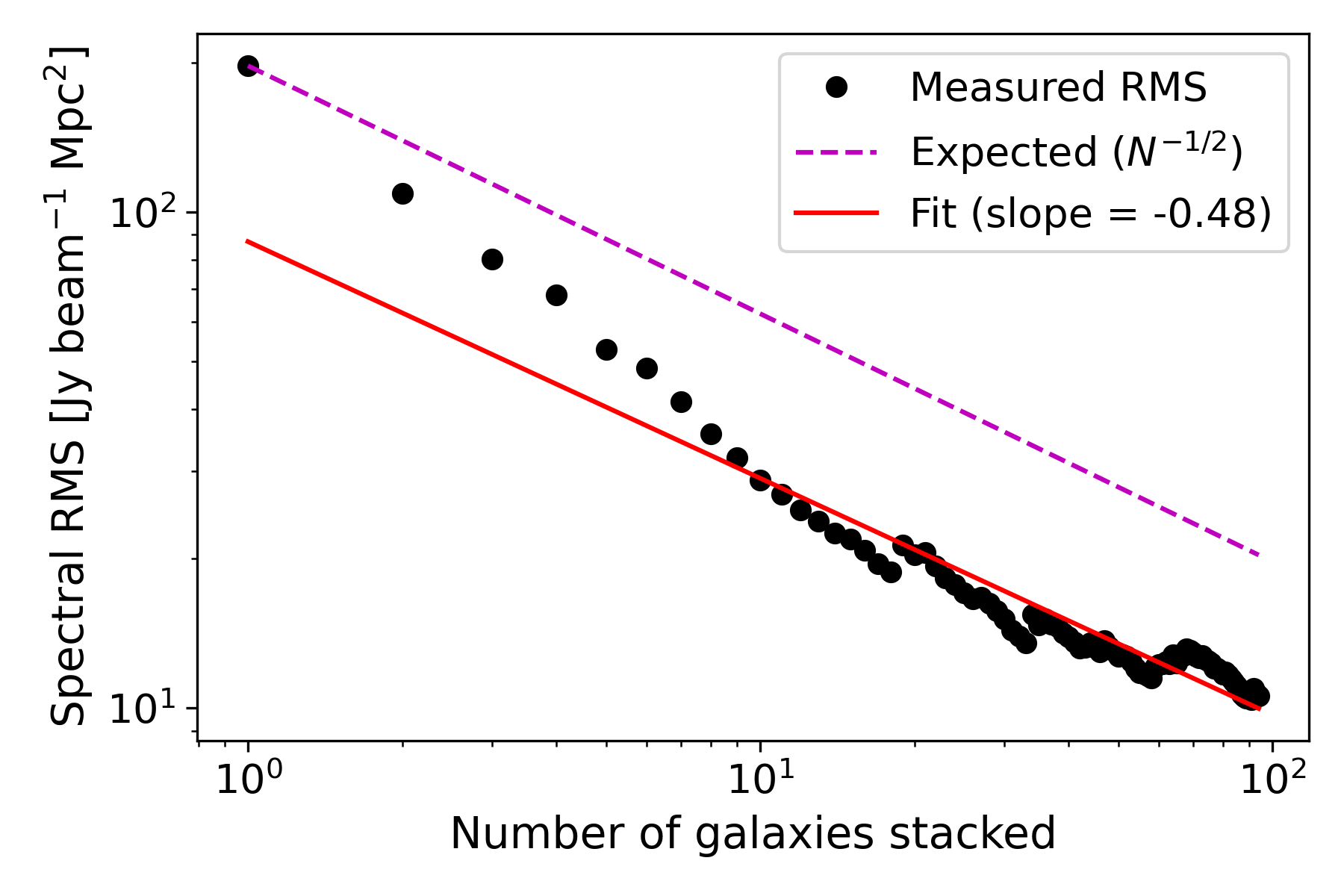}
\caption{The measured RMS noise of the stacked simulated spectrum as a function of the number of galaxies included in the stack, computed from line-free channels. The red line shows a least-squares fit to the data in log–log space. For ideal stacking, the noise is expected to scale as $N^{-1/2}$, corresponding to a slope of approximately $-0.5$, as shown by the dashed line. Our simulated noise scales down with a slope of $-0.48\pm0.02$. The slightly steeper decrease at low $N$ is likely due to small-number statistics and non-uniform noise properties in the sample, which average out as more galaxies are included in the stack.}
\label{fig:rms_sim}
\end{figure}

Because the \ion{H}{i} signal at these redshifts is extremely faint, only a small number of galaxies can be detected individually. To overcome this limitation, we use stacking techniques, which combine the spectra of many undetected galaxies to measure their average \ion{H}{I} content. This can only be done when the spectroscopic redshifts of sources that are likely to emit the 21\,cm line are known to a relatively high accuracy \citep{Maddox:2013}. The method used to do this is generally known as `spectrum stacking' \citep{Zwaan:2000,Chengalur:2001}. This involves the measurement of the spectrum of a sample of sources at the coordinates and redshifts where we expect them to be. To extract a collective signal, we align the spectra according to the galaxies’ known redshifts and combine them using a weighted average, thereby increasing the overall signal-to-noise ratio \citep{Delhaize:2013}. If the noise resembles a Gaussian distribution, it should reduce while the signal is co-added. This way, we can measure a spectrum that is representative of the entire sample of galaxies. If we observe a peak with a high enough integrated signal to noise ratio and a believable width, we have a stacked detection. This spectrum can then be analysed as any normal detection would with the only caveat being that any values calculated from it are now the mean of the sample. This technique was first shown to be successful in a study of the average properties of field galaxies by \citet{Lah:2007}. 

The method that we implement to stack LCBGs is a slightly altered approach known as `cubelet stacking' \citep{Chen:2021a}. Instead of extracting a spectrum around each individual source, a small region, or `cubelet', around the source is extracted. These cubelets are then aligned such that the right-ascension, declination and redshift of the center of each source is at the center. The cubelets are then stacked and a spectrum is extracted from the stacked cube. Along with the stacked sources, point spread function (PSF) cubelets are also stacked. The reason for using this approach as opposed to standard spectrum stacking is that, given the resolution of the CHILES data, it is more likely that sources will be resolved throughout the entire cube. Cubelet stacking results in a more accurate stack because we can perform a deconvolution of the stacked cube via a stacked PSF cube. This yields more accurate flux and mass measurements. The frequency-dependent change in the CHILES synthesised beam across a single cubelet is very small, because the cube was smoothed to a resolution of $9''\times9''$ across all channels. For this reason, it is not necessary to deconvolve each frequency slice with its own PSF prior to stacking. Constructing a stacked PSF cube is sufficient and captures the relevant frequency dependence at the level required for accurate flux recovery. Our method follows that as described in \citep{Luber:2025a}. Our stacking procedure was as follows:

\begin{enumerate}
    \item From an input source list with known coordinates and spectroscopic redshifts, we extract cubelets of $64\times64$ pixels and a width of 76 channels, or approximately 9.5\,MHz, with the source exactly at the center of the cubelet. These cubelets are then corrected to be centered at the \ion{H}{i} emission rest frequency.
    \item We calculate and apply the primary beam correction factor at the point of each source. 
    \item The emission in the cubelet is initially in units of Jy\,beam$^{-1}$. We convert this to luminosity density per beam (Jy\,beam$^{-1}$Mpc$^2$) using the distance of the galaxy in that cubelet calculated from its redshift.
    \item We stack our spectra using a weighted average. The weights for each cubelet were calculated as:
    \begin{equation}
        \label{eqn:weights}
        w_i=\frac{1}{\sigma_i^2}
    \end{equation}
    where $\sigma$ is the standard deviation of the cubelet measured at a region that falls outside the area of expected emission. This weighting scheme slightly increases the S/N of the stacked spectrum \citep{Delhaize:2013}. The cubes are then stacked as follows:
    \begin{equation}
        \label{eqn:stacking}
        \langle S\rangle=\frac{\sum_{i=1}^{n}{S_iw_i}}{\sum_{i=1}^{n}{w_i}}
    \end{equation}
     where $S_i$ represents a single cubelet and $w_i$ is its associated weight.
     \item The stacked cube is then cleaned using a stacked PSF cube by means of the H\"ogbom cleaning algorithm \citep{Hogbom:1974} down to 1.5 times the RMS noise.
     \item A spectrum is then extracted from the center of the stacked cubelet in a region of $14''\times14''$. This translates to a physical size of $\approx 50$\,kpc at $z\approx 0.22$ and $\approx 84$\,kpc at $z\approx 0.48$. Channels that show emission are identified. A 2D polynomial is fit to the off-emission channels and subtracted off. These baseline structures are typically associated with residual side-lobe patterns or imperfect continuum subtraction. They cannot be reliably modeled or removed before stacking if these components are well below the noise level in individual cubelets. A stacked detection is deemed successful if the rest frame emission is centered at 1420\,MHz with a high enough integrated signal to noise (S/N $>3$). We also inspect the noise around the integrated region by generating a moment 0 (\ion{H}{i} column density) map over the channels with emission.
     \item In order to make sure that our stacking process will be able to correctly search for a signal embedded in noise, we simulate \ion{H}{i} sources in a cube and perform our stacking technique on these sources.  We produce a cube of 760 channels and $640\times 640$ pixels, representing approximately $22'\times 22'$ in the sky and embedded them with 60 sources. We find that the co-added spectrum obtained from stacking behaves as expected, with the noise decreasing approximately as $N^{-1/2}$, where $N$ is the number of galaxies in the stack (see Figure~\ref{fig:rms_sim}).
\end{enumerate}

\section{Results}

\begin{figure*}
\includegraphics[width=0.9\textwidth]{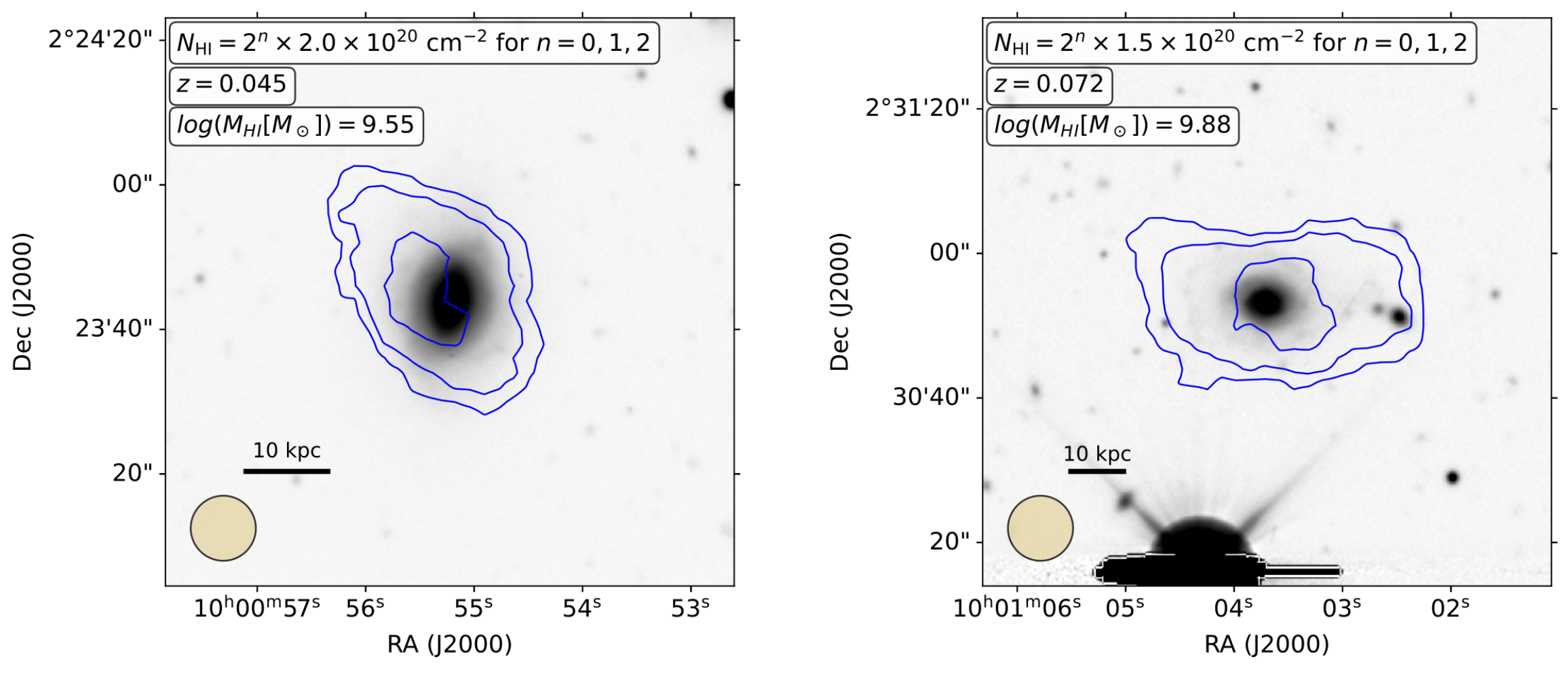}
\caption{\ion{H}{i} column density contours overlaid on optical DECaLS images. Left: The directly detected LCBG at $z=0.045$ with an estimated \ion{H}{i} mass of $(3.53\pm0.06)\times10^9\,M_\odot$. The spectrum of this detection had $W_{20}=(317\pm26)$\,km/s. Right: The directly detected LCBG at $z=0.072$ with an estimated \ion{H}{i} mass of $(7.55\pm0.23)\times10^9\,M_\odot$. The spectrum of this detection had $W_{20}=(211\pm26)$\,km/s.}
\label{fig:Direct_detections}
\end{figure*}

\begin{figure*}
\includegraphics[width=\textwidth]{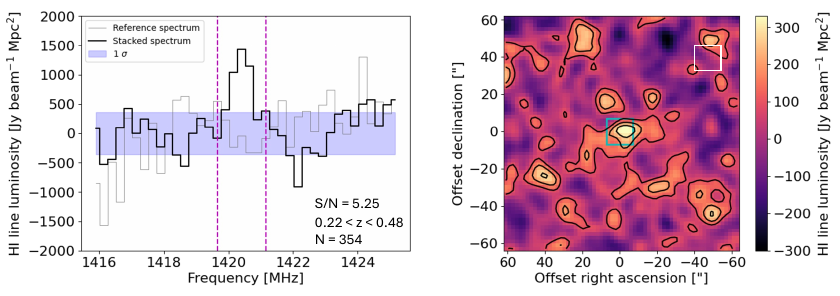}
\caption{The integrated, stacked spectrum from a stacked cubelet, smoothed to 50\,km/s for the LCBG sample between $0.22<z<0.48$. The spectrum is taken from the aperture indicated by the cyan box. The reference spectrum is taken from the aperture indicated by the white box. The filled purple region indicates the off-line RMS noise of the stacked spectrum. The moment 0 map on the right of each panel is produced from the channels between the rest-frame frequencies indicated by the dashed lines.}
\label{fig:Stacks_1}
\end{figure*}

\begin{figure*}
\includegraphics[width=\textwidth]{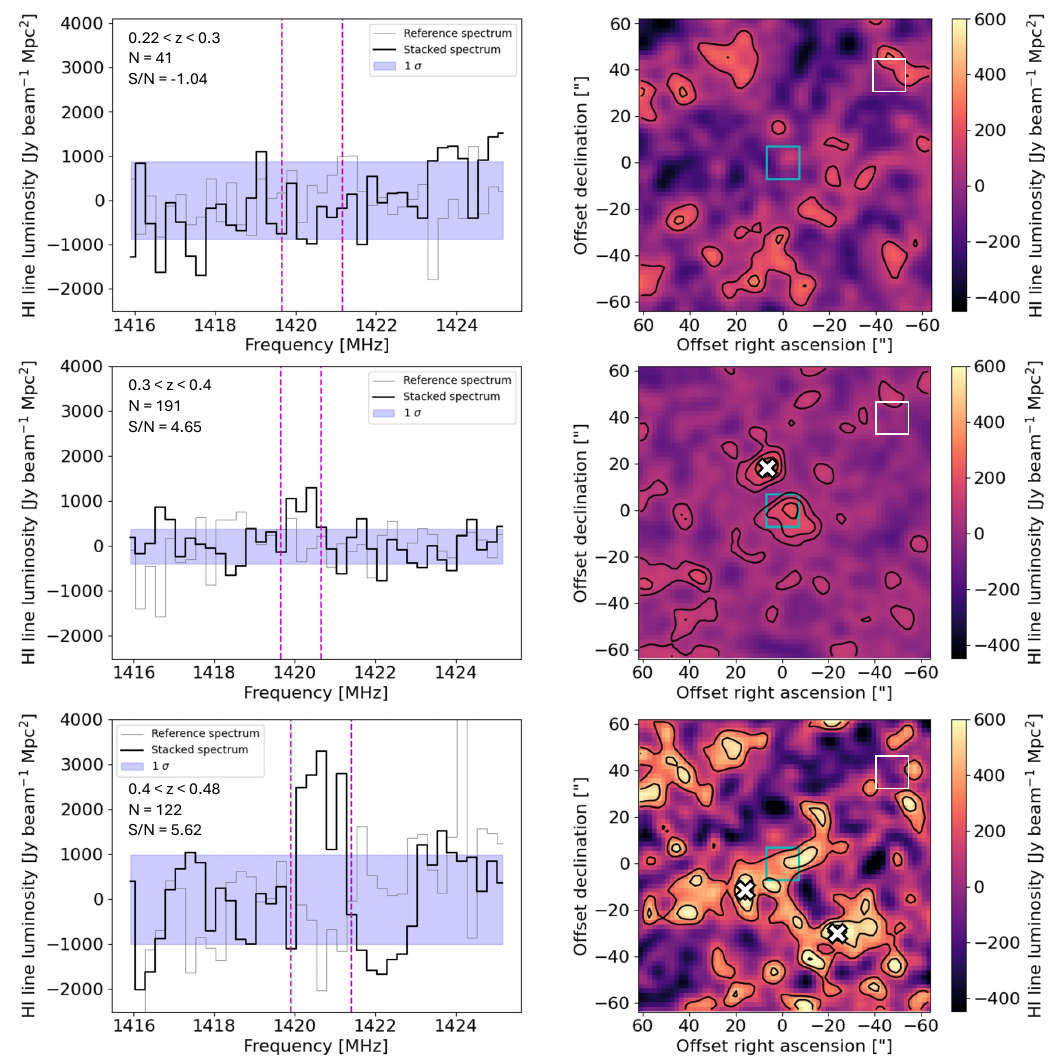}
\caption{The integrated, stacked spectrum from a stacked cubelet, smoothed to 50\,km/s for the LCBG sample between $0.22<z<0.3$ (top), $0.3<z<0.4$ (center) and $0.4<z<0.48$ (bottom). The spectrum is taken from the aperture indicated by the cyan box. The reference spectrum is taken from the aperture indicated by the white box. The filled purple region indicates the off-line RMS noise of the stacked spectrum. The moment 0 map on the right of each panel is produced from the channels between the dashed lines. The white crosses show the locations of notable noise peaks. We extract spectra at these points for further investigation to see how they compare to the cental detections.}
\label{fig:Stacks_5_6_7}
\end{figure*}

\begin{figure*}
\includegraphics[width=\textwidth]{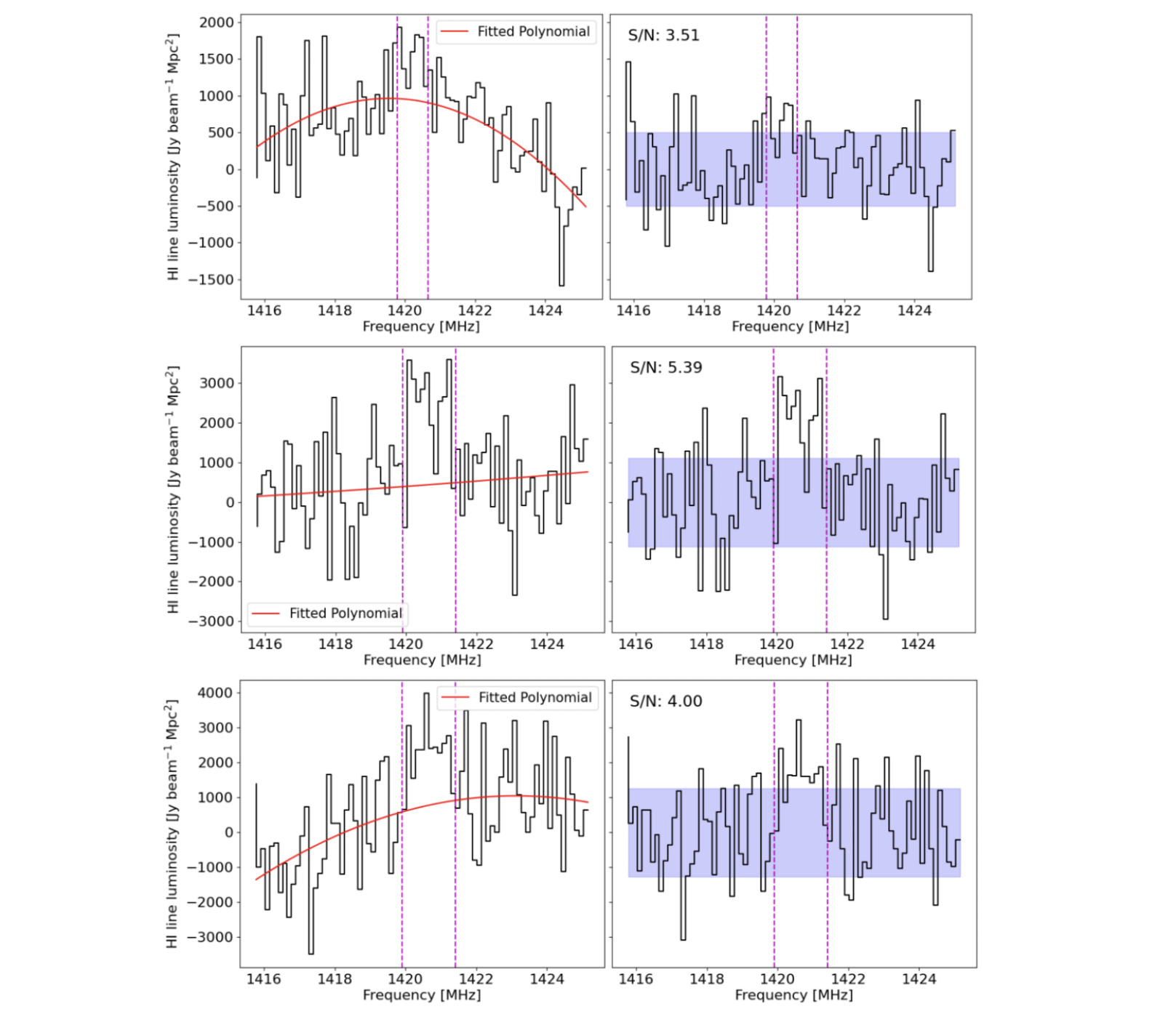}
\caption{Spectra taken at noise peaks present in some of our moment 0 maps indicated with white crosses in Figure~\ref{fig:Stacks_5_6_7}. Each spectrum is baseline corrected by subtracting a fitted 2nd order polynomial. The integrated signal to noise is calculated of the region spanning the same channels as that of the central peak in each stack. This value is shown in each plot. Top: Spectrum taken from the noise peak in the stack between $0.3<z<0.4$ located at (8'',18'') in the moment 0 map. Center: Spectrum taken from the noise peak in the stack between $0.4<z<0.48$ located at (18'',-15'') in the moment 0 map. Bottom: Spectrum taken from the noise peak in the stack between $0.4<z<0.48$ located at (-22'',-30'') in the moment 0 map.}
\label{fig:Noisepeaks}
\end{figure*}

\begin{figure*}
\includegraphics[width=\textwidth]{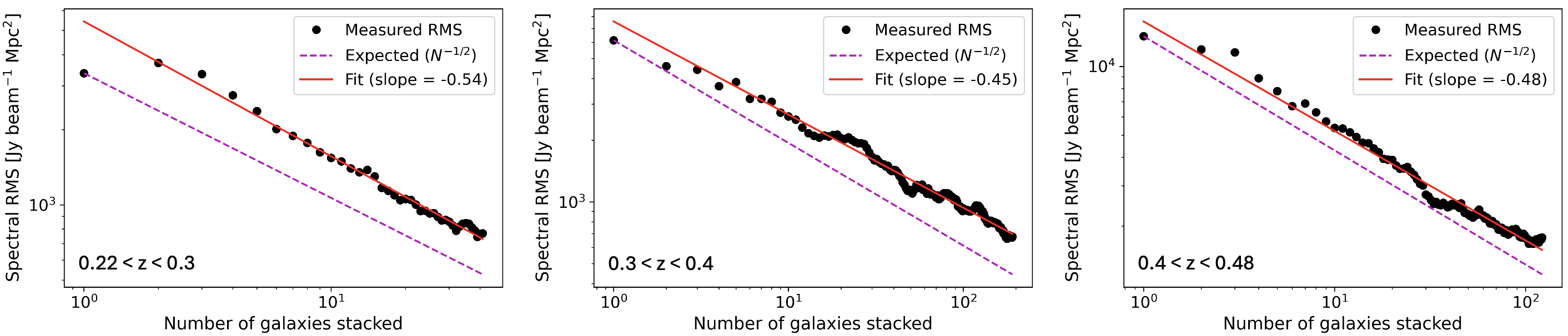}
\caption{We show the measured RMS noise of the stacked spectra as a function of the number of galaxies included in the stack, computed from line-free channels. The red line shows a least-squares fit in log–log space. The RMS follows the expected $N^{-1/2}$ scaling (the dashed line), with a fitted slopes of $-0.54\pm0.01$, $-0.45\pm0.01$ and $-0.48\pm0.01$ for the $0.22<z<0.3$, $0.3<z<0.4$ and $0.4<z<0.48$ stacks respectively. The offsets from the expected curve could be attributed to the noise of the original spectrum being abnormally high.}
\label{fig:stacked_rms}
\end{figure*}

\subsection{Direct detections}

\begin{table}
        \centering
        \caption{Measurements of the directly detected LCBGs.}
        \begin{tabular}{lcccccr}
              \hline
              $z$ &$\alpha$ &$\delta$& $D_{L}$ &$M_{\rm HI}$ & $W_{20}$  & $S/N$ \\
              & J2000 & J2000 & Mpc & $10^9 M_{\odot}$& km s$^{-1}$ & \\
              (1) & (2) & (3) & (4) & (5)& (6) & (7) \\
              \hline
              0.045 & 150.234 & 2.398 & 199.34 & $3.53 \pm 0.06$& $317\pm 26$& 23.25 \\
              0.072 & 150.265 & 2.514 & 325.13 &  $7.55 \pm 0.23$& $211\pm 26$& 13.50 \\
              \hline
       \end{tabular}
       {\raggedright \textit{Notes}. (1) \ion{H}{i} redshift; (2) right ascension; (3) declination; (4) luminosity distance; (5) average primary beam corrected \ion{H}{i} mass with uncertainties propagated from the integrated flux density; (6) linewidth with the error set by the channel width; (7) integrated signal to noise ratio.\par}
       \label{tab:DirectDetections}
\end{table}

We reanalyse two directly detected LCBGs which were previously studied by \citet{Hunt:Thesis} using the first epoch of CHILES data. Each detection's \ion{H}{i} mass is measured. All measured properties are shown in Table \ref{tab:DirectDetections}. We show column density maps of these two detections overlaid on optical DECaLS images in Figure \ref{fig:Direct_detections}. Looking at the first detection on the left, which lies at $z\approx 0.045$, the \ion{H}{i} column density contours correspond closely with the optical counterpart of the LCBG. We could be observing some disruption in the gas with respect to the stellar component towards the northern part of the image. We measure its \ion{H}{i} mass to be $(3.53\pm0.06)\times10^9\,M_\odot$. The \ion{H}{i} emission has a semi-major axis of $40.4''$ or $\approx 35.75$\,kpc. The emission has a linewidth of $W_{20}=317\pm 26$\,km s$^{-1}$. The stellar mass and star formation rate for this galaxy is not known. In the second detection at $z\approx0.072$ we again see that the \ion{H}{i} contours surround the stellar component with the major axes approximately aligned. Its \ion{H}{i} mass is found to be $(7.55\pm0.23)\times10^9\,M_\odot$ with the \ion{H}{i} emission spanning a semi-major axis of $\approx39''$ or $\approx 53.49$\,kpc. Its emission has a linewidth of $W_{20}=211\pm 26$\,km s$^{-1}$. The galaxy has a stellar mass of $\log (M_* [M_{\odot}])=10.179$ and a SFR of 5.41\,$M_{\odot}$\,yr$^{-1}$ \citep{Laigle:2016}. We note that both detections show a misalignment of less than one beam width between the optical and radio counterparts.

\subsection{Stacked results}
With only two direct detections, we can estimate the average \ion{H}{i} mass of LCBGs by stacking the non-detections. In order to do this, we first have to decide on the optimal bins in which to group LCBGs. If we look at Figure~\ref{fig:CHILES_noise}, we coincidentally observe the number of LCBGs drop at around the same frequency range where the noise in the cube increases ($940\lesssim \nu \lesssim 1160$). The ideal bin contains a large enough sample of galaxies to stack for it to be statistically significant, while also avoiding channels which contain a large amount of noise. We follow the following procedure after performing a stack: We extract an integrated spectrum from the central square region of the stacked cubelet with a fixed size of $14''\times 14''$ and then smooth the spectrum to a resolution of $\approx 50$\,km/s for a higher S/N. A moment 0 map is then produced over the channels which show emission. If we do not observe a spectral detection, we still continue to produce a moment map over channels spanning a velocity width of 227\,km/s centered at the rest frequency of the stacked galaxies. This is the average $W_{20}$ line width measured from a sample of local LCBGs by \citet{Garland:2004} and would allow us to still inspect the moment map for further analysis. Finally we would inspect both the spectrum and the moment map to see whether we observe a detection both spectrally and spatially.

We first perform one large stack of 354 LCBGs between $940<\nu<1160$\,MHz ($0.22<z<0.48$). We omit galaxies with redshifts less than 0.22, because, as can be seen in Figure \ref{fig:CHILES_noise}, the RMS noise increases dramatically above this frequency, while the number of LCBGs starts to get very low. This stack, shown in the left panel of Figure~\ref{fig:Stacks_1}, results in a detection with an integrated signal to noise ratio of 5.25. The right panel shows a clearly visible detection in the center of the stacked cubelet. Clearly we have detected neutral hydrogen in this redshift range in the CHILES cube. From here we continue to calculate the average \ion{H}{i} masses of the galaxies in our stacking using:
\begin{equation}
    \frac{M_{\rm HI}}{M_{\odot}}=49.8\left(\frac{D_L}{\textrm{Mpc}}\right)^2\int\frac{S(\nu)}{\textrm{Jy}}\left(\frac{d\nu}{\textrm{Hz}}\right)
    \label{eqn:HI_mass}
\end{equation}
Here $\int D_L^2S(\nu)d\nu$ is the flux from our stacked spectrum integrated over the channels showing emission, multiplied by the square of the average luminosity distance of the galaxies in our stack. All calculated masses are listed in Table \ref{tab:Stacked_properties}. Our primary goal was to trace the evolution of \ion{H}{i} across redshift, so we split this range into three bins of equal frequency width. We decided to bin in terms of frequency, since that is how the cube's channels are defined. The bins are $940<\nu<1013$\,MHz (122 galaxies between $0.4<z<0.48$), $1013<\nu<1086$\,MHz (191 galaxies between $0.3<z<0.4$) and $1086<\nu<1160$\,MHz (41 galaxies between $0.22<z<0.3$). These stacks are shown in Figure~\ref{fig:Stacks_5_6_7}. We start losing detections when we stack certain bins, indicating a lack of neutral hydrogen in some of the subsets of LCBGs or too much of an increase in noise. We can still measure a upper limit for the \ion{H}{i} mass, $\sigma_{\ion{H}{i}}$, from the non-detections. From Equation \ref{eqn:HI_mass} we can derive:
\begin{equation}
    \sigma_{\rm HI}=49.8D_L^2\sigma_Sd\nu\sigma_{S/N}
\end{equation}
Here, $D_L^2\sigma_S$ represents the error in the cubelet in luminosity density units. We calculate this by resampling regions within the cubelet with aperture sizes comparable to that in which we would extract the stacked spectrum. $d \nu$ is the width of our stacked cubelet's frequency channels. We set $\sigma_{S/N}$ (our signal to noise ratio) to 3 to give us $3\sigma$ upper limit estimates of the mass. This is calculated over a velocity width of 227\,km/s, which is the average $W_{20}$ linewidth measured from a sample of local LCBGs by \citet{Garland:2004}.

\begin{table}
    \centering
    \caption{The properties of the individual stacks.}
    \label{tab:Stacked_properties}
    \begin{tabular}{lccccr}
        \hline
        Figure & $z$ range & $\nu$ range& N & $\langle M_{\rm HI} \rangle$ & $S/N$\\
        & &(MHz)& &($\times 10^9 M_{\odot}$)\\
        (1) & (2) & (3)& (4) & (5) & (6)\\
        \hline
        \ref{fig:Stacks_1} & 0.22-0.48 & 940-1160 & 354 & $ 3.42 \pm 1.10$ & 5.25 \\
        \ref{fig:Stacks_5_6_7} (top)& 0.22-0.3 & 1086-1160 & 41 & $<4.89$ & -1.04\\
        \ref{fig:Stacks_5_6_7} (center)& 0.3-0.4 & 1013-1086 & 191 & $2.49 \pm 0.75$ & 4.65\\
        \ref{fig:Stacks_5_6_7} (bottom)& 0.4-0.48 & 940-1013 & 122 & $6.44 \pm 2.71$ & 5.62\\
 	\hline
    \end{tabular}
    {\raggedright \textit{Notes}. (1) Figure number in this paper; (2) redshift range; (3) frequency range; (4) number of galaxies in the stack; (5) average \ion{H}{i} mass; (6) integrated signal to noise ratio.\par}
\end{table}

We see no detection in the lowest redshift stack (containing 41 galaxies). We can only gather an upper limit for the mass: $\langle M_{\rm HI} \rangle < 4.89$\,$\times 10^9 M_{\odot}$. In the central bin (191 galaxies) we observe a detection in the spectrum. We can make out a slight detection in the center of the moment map (see center panel of Figure~\ref{fig:Stacks_5_6_7}) with S/N$=4.65$. This is accompanied by what seems to be a noise peak i.e. a bright spot in the moment map caused by abnormally high noise in some of the channels over which we are integrating. The top panel of Figure~\ref{fig:Noisepeaks} shows our inspection of this noise peak in which we measure a spectrum at its location. We observe quite a large baseline offset, which we correct by fitting and subtracting a 2nd order polynomial. We measure the integrated S/N of this corrected spectrum over the same channel range as we did for the central emission. This comes to 3.51. The fact that the noise peak has a lower S/N than that of the central detection (4.65) and the presence of the large baseline offset, convinces us that our central detection is indeed significant and real. The highest of the redshift stacks (122 galaxies) shows a potential non-localised detection (S/N$=5.62$). Again we observe many other bright regions in the map. In the center and bottom panels of Figure~\ref{fig:Noisepeaks} we show our analysis of these features. Both of these show signal to noise ratios which are less than, but still comparable, to the central emission. Because the S/N of the central emission is higher and is peaked both spatially and spectrally, we still continue to measure the \ion{H}{i} mass over the indicated channels in bottom panel of Figure~\ref{fig:Stacks_5_6_7}. The other peaks seem to be consistent with baseline fluctuations or artifacts within the cube, as seen in Figure \ref{fig:Noisepeaks}. We note that this frequency range of the cube is prone to elevated kurtosis values \citep{Luber:2025a}, which may cause the noise to deviate from perfect Gaussianity. 

We further investigate the behavior of the noise in our stacks by calculating the off-line RMS of our stacked spectra as we add galaxies to the stack. If the noise in the cube is purely Gaussian, we would expect the RMS to scale down by $N^{-1/2}$, where $N$ is the number of galaxies included in the stack. As can be seen in Figure \ref{fig:stacked_rms}, the RMS does scale as expected, but does tend to be systematically higher across all redshift bins. This does however prove that our stacks are scaling correctly and that deviating RMS noise is caused by non-Guassianity in certain parts of the data cube.

\begin{figure*}
\includegraphics[width=\textwidth]{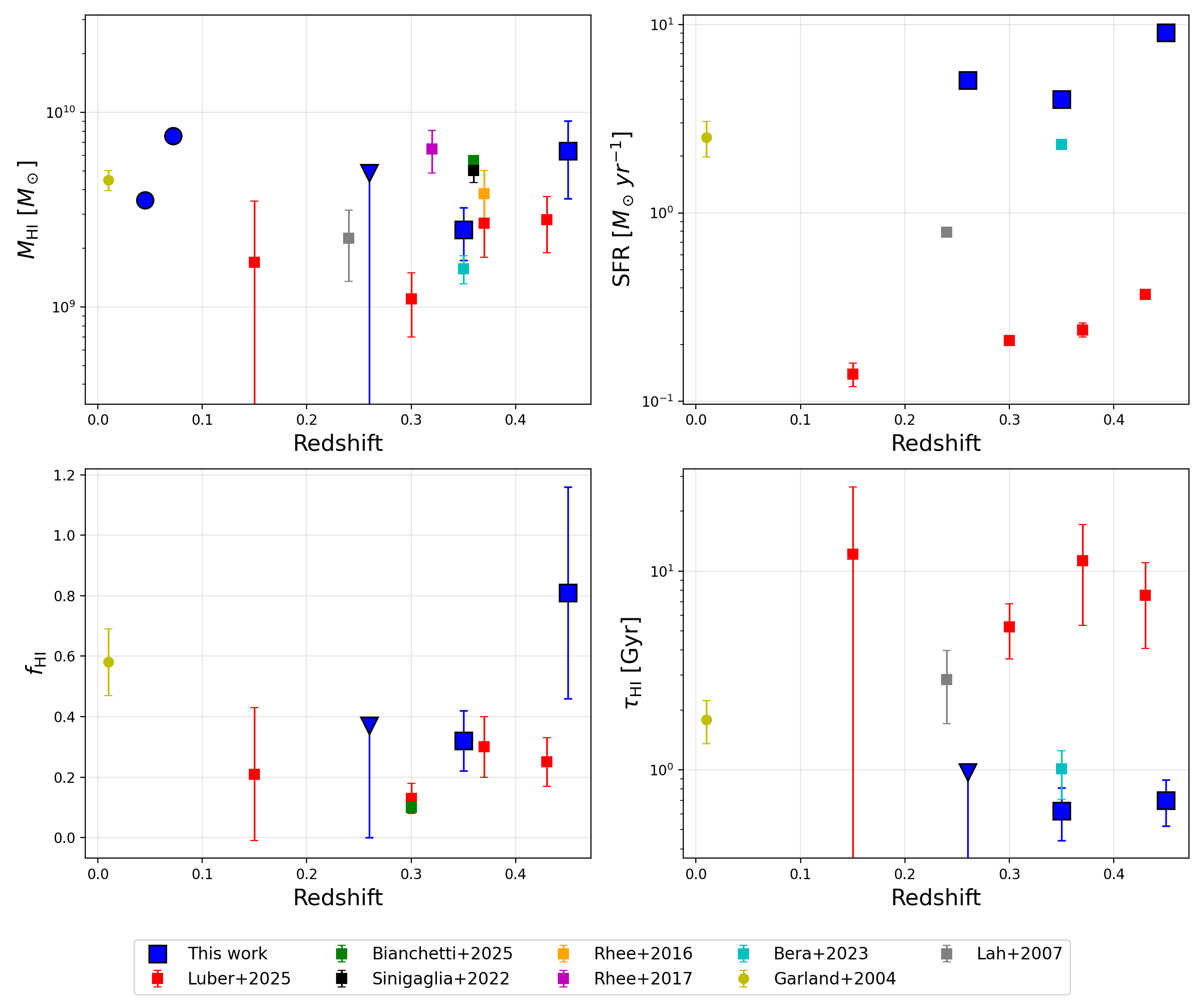}
\caption{Evolution of LCBG properties compared to literature samples. Top left: average \ion{H}{i} masses as a function of redshift. Top right: star formation rates (SFRs) calculated from the COSMOS2015 catalogue \citep{Laigle:2016}. Note that error bars are smaller than the symbol size where not visible. Bottom left: \ion{H}{i} gas fractions, $f_{\rm HI} = M_{\rm HI}/M_\ast$. Bottom right: gas depletion timescales, $\tau_{\rm HI} = M_{\rm HI}/\mathrm{SFR}$. Blue markers show results from this work. Squares represent results from stacked averages. Circles represent direct detections or averages thereof. Triangles represent upper limit values.}
\label{fig:Discussion_plots}
\end{figure*}

\section{Discussion}

In the top left of Figure~\ref{fig:Discussion_plots} we plot our results from the stacks split over three bins. We also include the estimated masses of our two direct detections.  We do not see any clear evolution of \ion{H}{i} mass with redshift. To place our results in context, we compare them to stacked \ion{H}{i} masses of blue, star-forming galaxies at similar redshifts \citep{Bianchetti:2025,Sinigaglia:2022,Rhee:2018,Bera:2023,Lah:2007}, as well as to other starburst studies \citep{Rhee:2016}, and find good agreement across these samples. \citet{Luber:2025a} also did not see much evolution in \ion{H}{i} mass in their subset of blue, star-forming galaxies with masses comparable to that of our own ($10^{9-10}$\,M$_\odot$). Likewise, we do not observe a decline in \ion{H}{i} mass when considering our direct detections at lower redshift. This trend is consistent with the average \ion{H}{i} mass of directly detected low-redshift LCBGs reported by \citet{Garland:2004} using single dish observations. Since LCBGs retain their gas, they are likely evolving into gas-rich systems such as disc galaxies, alongside other, less intensely star-forming populations. 

Average star formation rates for the galaxies in our stacks in Figure \ref{fig:Stacks_5_6_7} were taken and calculated from the COSMOS2015 catalogue \citep{Laigle:2016}. Their SFRs were based on template fitting. We show these in the top right of Figure~\ref{fig:Discussion_plots}. As expected for starbursts like LCBGs, the galaxies in our sample exhibit elevated star formation rates (SFR$\approx6$\,M$_\odot$yr$^{-1}$) across $0.22<z<0.48$ compared to other blue, star-forming galaxies. While other galaxies in the CHILES study by \citet{Luber:2025a} show a decline in SFR toward lower redshifts with an average of $\approx0.24$\,M$_\odot$yr$^{-1}$, the LCBGs in our sample sustain a consistent rate of star formation across this range, which is to be expected, since they were selected based on their star-bursting characteristics. For comparison at higher $z$, we also plot the average SFR from the sample in \citet{Lah:2007}, calculated from H$\alpha$ line luminosity.  Finally we show the value derived from the SFR–$M_\ast$ relation of \citet{Bera:2023}, which again show the trend of lower SFRs toward lower redshifts.
At $z\sim0$, the average SFR of directly detected LCBGs from \citet{Garland:2004} is lower than that of our stacked LCBGs at higher redshifts, suggesting a slight decrease in star formation efficiency at the lowest redshifts. Overall, the SFR of LCBGs appears to decline more slowly than that of other star-forming galaxies.

The \ion{H}{i} gas fraction, defined as $f_{\rm HI} \equiv M_{\rm HI}/M_\ast$ and shown in the bottom left of Figure~\ref{fig:Discussion_plots}, provides a measure of how much fuel a galaxy has relative to its stellar content. For LCBGs, we again see no clear declining trend with redshift; they appear to retain substantial gas reservoirs even at low redshifts. It is important to note that \citet{Garland:2004} report only dynamical masses calculated within the effective radii, $M_{\rm dyn}(R<R_e)$, and within the isophotal radii, $M_{\rm dyn}(R<R_{25}(B))$. For our comparisons we adopt the former, since $M_{\rm dyn}(R<R_{25}(B))$ would also include dark matter. Overall, LCBGs appear comparable in \ion{H}{i}-richness when compared the broader star-forming population, yet they do not seem to exhaust their gas as rapidly.

Another measure of how efficiently the \ion{H}{i} gas in LCBGs is converted into stars is the gas depletion timescale, defined as $\tau_{\rm HI} \equiv M_{\rm HI}/\mathrm{SFR}$ and shown in the bottom right of Figure~\ref{fig:Discussion_plots}. Our LCBGs consistently show very short depletion timescales compared to other blue galaxies. This indicates that they are in an intense star-forming phase and are likely to evolve into different galaxy types once their gas is consumed or stabilised. Our \ion{H}{i} observations corroborate the fact that LCBGs rapidly decline in number density with time. We calculate $\tau_{\rm HI}$ from the $\tau_{\rm HI}$–$M_\ast$ relation at $z\sim0.35$ presented in \citet{Bera:2023}, using the average stellar mass of our LCBGs at that redshift ($\tau_{HI}\approx1.01\pm0.3$)\,Gyr and find good agreement with our value. This is somewhat surprising, given that the sample is not composed exclusively of starburst systems, but rather of blue, star-forming galaxies confirmed to be free of AGN activity. A plausible explanation is that a substantial fraction of the sample may nonetheless be undergoing starburst phases. The timescale reported by \citet{Garland:2004} is also low, suggesting that LCBGs remain efficient star formers even at low redshifts. The key result from these short depletion timescales, taken together with the high gas fractions, is that LCBGs maintain large gas reservoirs but consume them rapidly, consistent with being in a burst or transitional phase.

Together, the trends in SFR, $f_{\rm HI}$, and $\tau_{\rm HI}$ suggest that LCBGs occupy a distinct evolutionary pathway compared to the broader star-forming population. They combine elevated star formation rates with substantial \ion{H}{i} reservoirs, yet consume this gas on very short timescales. This behavior implies that LCBGs are in a highly efficient phase of growth, during which they rapidly build up their stellar mass while still retaining large gas supplies. Rather than fading quickly into quiescent systems, the evidence shows their eventual evolution into gas-rich disc galaxies or bulge and disc systems. This idea is strengthened by the fact that most of the LCBGs in the local sample in \citet{Garland:2004} were spirals, barred spirals or irregular. A subsample of these local LCBGs were studied by \citet{Rabidoux:2018} and they found that they tend to be rotationally supported and have nearby companions. LCBGs represent a key transitional population, bridging actively star-forming galaxies and the more settled disc or irregular systems as seen at lower redshifts, depending on their interactions with other galaxies over time. 

\section{Summary and Conclusions}

We have presented the first measurements of the average \ion{H}{i} content of LCBGs over the redshift range $0<z<0.48$ using the full CHILES dataset. By using a cubelet stacking technique, we were able to detect the average \ion{H}{i} emission of LCBGs at intermediate redshifts, despite the faintness of individual detections. Our results indicate that the average \ion{H}{i} masses of LCBGs remain roughly constant over this redshift range, with no clear evidence for strong evolution. 
Alongside the stacked detections, we also report two direct \ion{H}{i} detections of LCBGs at low redshift, with masses consistent with those measured locally in previous studies. Given the intrinsic spread in \ion{H}{i} masses observed among LCBGs, our detections likely represent the \ion{H}{i}-rich end of the population, while the non-detections remain consistent with typical LCBG \ion{H}{i} masses falling below our detection threshold. As such, the detected galaxies do not appear anomalous. Comparisons with local LCBGs suggest that the gas reservoirs of these galaxies have not significantly declined between $z\sim 0.48$ and the present day. However, the size of our sample and the presence of artifacts in the stacked moment maps introduce uncertainties in our results.
Our analysis further shows that LCBGs can be differentiated from the broader star-forming population by their ability to sustain high gas fractions and elevated star formation rates across $0<z<0.48$. While typical star-forming galaxies often exhibit a steady decline in both gas content and SFR with cosmic time, LCBGs maintain substantial \ion{H}{i} reservoirs and vigorous star formation activity well into intermediate redshifts ($z\sim0.48$). Their short gas depletion timescales indicate that they are consuming their gas with high efficiency, consistent with being in a transitional phase of evolution. 
Our analysis suggests that LCBGs are thus not likely to evolve into lower mass quenched elliptical galaxies, but rather spirals or barred spirals. If they were to experience quenching, it is more likely that it would be due to interactions with companions, as opposed to internal processes.

To further constrain the evolution of LCBGs, future work should focus on expanding the available samples and improving observational sensitivity. A larger local sample would provide more robust measurements of \ion{H}{i} and stellar masses, offering a clearer benchmark for the low-redshift population. Deeper \ion{H}{i} observations such as forthcoming data from the Looking At the Distant Universe with the MeerKAT Array (LADUMA) survey \citep{Blyth:2016} will both increase the number of direct detections and enhance the quality of stacked spectra by lowering the noise per channel and boosting the statistical significance of average trends. It would also increase the redshift range over which they can be found. Ultimately, this would allow us to place tighter constraints on the evolutionary fate of these galaxies.

\section*{Acknowledgements}
We thank the anonymous referee for the constructive comments that improved this paper. Henco Arlow and D.J. Pisano acknowledge support by the South African Research Chairs Initiative (SARChI) of the Department of Science and Technology and National Research Foundation. We acknowledge the use of the ilifu cloud computing facility – \href{www.ilifu.ac.za}{www.ilifu.ac.za}, a partnership between the University of Cape Town, the University of the Western Cape, Stellenbosch University, Sol Plaatje University and the Cape Peninsula University of Technology. The ilifu facility is supported by contributions from the Inter-University Institute for Data Intensive Astronomy (IDIA – a partnership between the University of Cape Town, the University of Pretoria and the University of the Western Cape), the Computational Biology division at UCT and the Data Intensive Research Initiative of South Africa (DIRISA). This work made use of the CARTA (Cube Analysis and Rendering Tool for Astronomy) software (\href{https://zenodo.org/records/17050846}{DOI 10.5281/zenodo.3377984} – \href{https://cartavis.github.io}{https://cartavis.github.io}). The National Radio Astronomy Observatory is a facility of the National Science Foundation operated under cooperative agreement by Associated Universities, Inc.

\section*{Data Availability}
The CHILES data cube is described in van Gorkom et al. (in preparation). The Jupyter notebooks used to perform the stacking analysis will be made available upon reasonable request.


\bibliographystyle{mnras}
\bibliography{biblio} 





\bsp	
\label{lastpage}
\end{document}